\pdfoutput=1
\documentclass[11pt]{article}

\usepackage[preprint]{acl}

\usepackage{times}
\usepackage{latexsym}
\usepackage{url}
\usepackage{graphicx}
\usepackage{booktabs}
\usepackage{multirow}
\usepackage{graphicx}
\PassOptionsToPackage{table,xcdraw,dvipsnames}{xcolor}
\usepackage{amsmath}
\usepackage{mathrsfs}
\usepackage{amssymb}
\usepackage{amsthm}   % 引入定理、引理等环境
\theoremstyle{definition} % 设置定理样式为定义样式
\theoremstyle{plain} % 设置字体样式为斜体，默认样式为斜体
\usepackage{algorithm}
\usepackage{algpseudocode}
\usepackage{booktabs}
\usepackage{caption}
\usepackage[dvipsnames]{xcolor}
\usepackage{hyperref}
\usepackage{booktabs}
\usepackage{makecell}
\usepackage{colortbl}  % 已经有了就不用管

\title{FIER: Fine-Grained and Efficient KV Cache Retrieval for Long-context LLM Inference}

\author{
  \textbf{Dongwei Wang\textsuperscript{1}},
  \textbf{Zijie Liu\textsuperscript{2}},
  \textbf{Song Wang\textsuperscript{3}},
  \textbf{Yuxin Ren\textsuperscript{1}},
  \textbf{Jianing Deng\textsuperscript{4}},
\\
  \textbf{Jingtong Hu\textsuperscript{4}},
  \textbf{Tianlong Chen\textsuperscript{2}},
  \textbf{Huanrui Yang\textsuperscript{1$\dagger$}},
\\
  \textsuperscript{1}The University of Arizona,
  \textsuperscript{2}The University of North Carolina at Chapel Hill,
\\
  \textsuperscript{3}The University of Virginia,
  \textsuperscript{4}University of Pittsburgh
\\
\small
\texttt{ \{dongweiw, yuxinr, huanruiyang\}@arizona.edu } \\
\small
\texttt{ \{jesseliu, tianlong\}@cs.unc.edu}, \texttt{ sw3wv@virginia.edu } \\
\small
\texttt{ \{jthu, JID70\}@pitt.edu }
}

\begin{document}
\maketitle
\renewcommand{\thefootnote}{\fnsymbol{footnote}} % 使用符号编号
\footnotetext[2]{Corresponding author}           % 2 对应 †
\renewcommand{\thefootnote}{\arabic{footnote}}  % 恢复数字编号
\begin{abstract}
The Key-Value (KV) cache reading latency increases significantly with context lengths, hindering the efficiency of long-context LLM inference. To address this, previous works propose retaining a small fraction of KV cache based on token importance. For example, KV eviction uses static heuristics to retain tokens, while KV retrieval dynamically selects query-relevant tokens for more adaptive cache management. However, we observe that important tokens are often sparsely distributed across the long context. This sparsity makes existing page-level KV retrieval inaccurate, as each page may include irrelevant tokens and miss critical ones. In this work, we propose Fier, a \underline{Fi}ne-Grained and \underline{E}fficient KV cache \underline{R}etrieval method. Fier uses 1-bit quantized keys to estimate the importance of each token, resulting in efficient and precise retrieval. Experiments show that Fier matches full KV performance using only 11\% of the cache budget across various long-context tasks, reducing decoding latency by 1.2$\times$ to 1.5$\times$. Code is available at \href{https://github.com/SimWangArizona/FIER}{here}.
\end{abstract}

\section{Introduction}
% Large Language Models (LLMs) have transformed the landscape of natural language processing. A key factor behind their success lies in their strong capability to understand and reason over long context. Recent models have pushed context lengths from 2k tokens in GPT-2 \cite{radford2019language} to 1M \cite{yang2025qwen2}, and even extending to 10M tokens \cite{team2024gemini}.

KV caching is a memory-for-computation acceleration technique for LLM inference, enabling faster decoding by reusing intermediate hidden states \cite{waddington2013kv}. 
However, during inference, each decoded token must attend to the full KV cache, causing cache reading latency to grow significantly with context length. For example, in LLaMA 7B \cite{touvron2023llama}, a 32k-token KV cache requires 16GB of memory and over 11ms to read—accounting for more than half of the total inference time \cite{tang2406quest}.

\begin{figure}[t]
    \centering
    \includegraphics[width=1\linewidth]{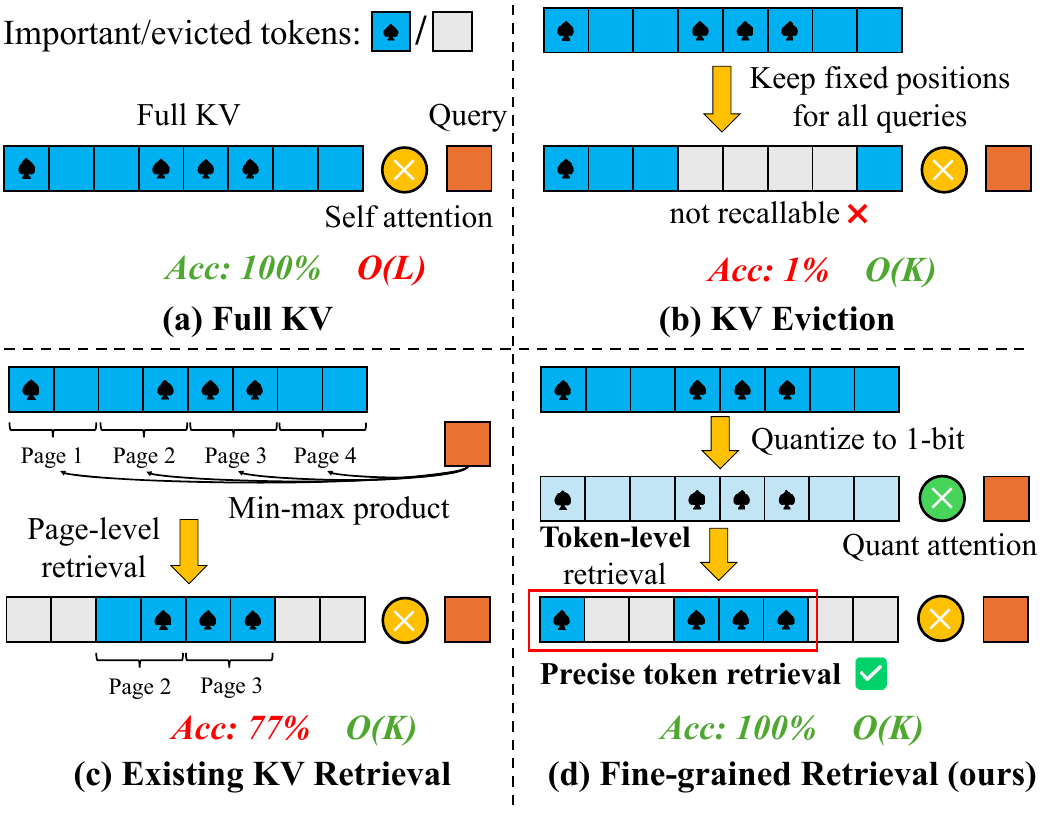}
    \vspace{-19pt}
    \caption{Comparison of KV eviction (b), KV retrieval (c) and Fier (d). While existing retrieval methods suffer from coarse granularity, Fier achieves higher accuracy through fine-grained \textit{token-level} retrieval, and preserves selection efficiency by quantization.}
    \label{intro1}
    \vspace{-16pt}
\end{figure}

To address this issue, previous works have proposed to selectively retain only a subset of KV cache entries based on token importance. Among them, one line of work—known as KV eviction (Fig.~\ref{intro1}(b))—focuses on retaining fixed-position tokens that typically receive higher attention weights, such as the initial tokens (due to the attention sink phenomenon) and the most recent tokens \cite{xiao2023efficient, zhang2023h2o, li2024snapkv, liu2023scissorhands}. However, these approaches overlook the dynamic nature of KV criticality, i.e., tokens that are evicted earlier may become important in the future. Their inability to recall evicted tokens often results in degraded performance in multi-round QA or long conversation applications. Motivated by this, another line of work—KV retrieval (Fig.~\ref{intro1}(c))—has been proposed to dynamically recall tokens that are most relevant to the current query during generation \cite{tang2406quest, chen2024arkvale}.

Despite achieving better performance, KV retrieval requires frequent estimation of the token importance for every new query, introducing additional computational overhead. To mitigate this, existing methods perform page-level retrieval, where all tokens that belong to a certain page of the KV cache are retrieved (or not retrieved) simultaneously to avoid computing full attention scores, leading to a coarser granularity of retrieval. 

%are grouped into coarse-grained  grouping tokens into segments and retrieving them at a coarser granularity to avoid computing full attention scores.
 
However, we observe that in long-context tasks, information relevant to the current query may be scattered throughout the entire input, i.e., important tokens are often sparsely distributed across the KV cache (shown in Fig.~\ref{ob1}).  Consequently, page-level retrieval inevitably leads to imprecise selection: \textit{retrieved pages may include irrelevant tokens, while evicted pages may exclude critical ones}, thereby affecting the model performance.

In this paper, we aim to address the retrieval inaccuracy while preserving selection efficiency. Specifically, we find that quantizing the key cache to as low as 1-bit has minimal impact on the accuracy of Top-$k$ selection in token importance estimation (shown in Fig.~\ref{ob2}). Despite quantization truncates large values, important tokens are still preserved in the Top-$k$ after computing quantized attention. Based on this insight, we propose \underline{Fi}ne-Grained and \underline{E}fficient KV cache \underline{R}etrieval (Fier), a 1-bit quantization-based KV retrieval method. As shown in Fig.~\ref{intro1}(d), Fier enables more accurate recovery of important tokens and reduces the selection cost. This leads to improved model performance under the same cache budget.

We evaluate Fier across PG19 \cite{rae2019compressive}, LongBench \cite{bai2023longbench}, and the passkey retrieval benchmarks \cite{peng2023yarn}. The results demonstrate the effectiveness of Fier in both generative and retrieval-focused settings. Experiments show that Fier achieves performance comparable to using the full KV cache while requiring only 11\% of the cache budget, and consistently outperforms existing KV eviction and retrieval baselines. Additionally, Fier achieves 1.2$\times$ to 1.5$\times$ decoding speedup across different context lengths on a single RTX 4090 GPU.
In summary, we make the following contributions in this work:
\begin{itemize}
    \item We observe the sparse distribution of important tokens within the KV cache in long-context scenarios, highlighting the necessity of fine-grained retrieval.
    \item We propose Fier, a KV retrieval method built on 1-bit key quantization, which enables efficient and accurate token-level retrieval.
    \item We conduct comprehensive evaluations of Fier across diverse long-context tasks and model architectures, demonstrating its superior performance and efficiency. 
\end{itemize}
\section{Related Work}
\textbf{Long-Context LLMs.} 
Large Language Models (LLMs) have transformed the landscape of natural language processing, largely due to their strong ability to deal with long context. Their context length capacity has increased dramatically—from 4k to 128k \cite{grattafiori2024llama}, 1M \cite{yang2025qwen2}, and even 10M \cite{team2024gemini} tokens. This expansion unlocks a range of advanced capabilities, including o1 long-range reasoning \cite{guo2025deepseek, openai2024reasoning}, in-context learning \cite{li2025minimax}, and multimodal intelligence \cite{weng2024longvlm}. Fier aims to improve the inference efficiency of long-context LLMs by exploiting the sparsity of the KV cache. 

\noindent \textbf{KV Cache Eviction.} Previous work identified the sparsity of attention matrices, showing that retaining only a small fraction of tokens is sufficient for the performance. For example, \citet{xiao2023efficient} propose to retain the first few tokens based on the “attention sink” phenomenon. H2O \cite{zhang2023h2o} retains a limited set of KV cache by selecting tokens with the highest cumulative attention scores. SnapKV \cite{li2024snapkv} selects clustered historical tokens along with a localized window of recent tokens. However, these approaches ignore the fact that tokens evicted can become important in the future. Fier addresses this via query-specific KV retrieval, enabling dynamic reassessment of token importance at each decoding step.

\noindent \textbf{KV Cache Retrieval.} KV retrieval methods, including our proposed Fier, dynamically select tokens relevant to the current query. However, existing approaches like Quest \cite{tang2406quest}, ArkVale \cite{chen2024arkvale}, PQCache \cite{zhang2025pqcache}, SparQattn \cite{ribar2023sparq} and MagicPIG \cite{chen2024magicpig} retrieve at the page or cluster level or compute partial tokens’ criticality for efficiency, overlooking the sparse distribution of important tokens. In this paper, we propose a fine-grained, token-level retrieval strategy that maintains efficiency while improving accuracy. This design better captures critical information missed by existing methods.

\noindent \textbf{KV Cache Quantization.} Another related line of work is KV quantization \cite{liu2024kivi, dong2024qaq}, which compresses the cache by performing self-attention in low-bit space. The objective of KV quantization is to minimize global reconstruction error. In contrast, Fier achieves cache compression by retaining only a subset of the full KV and adopts a relaxed quantization objective focused on preserving Top-ranked tokens, enabling the use of extremely low bit-widths.

\section{Methodology}
\begin{figure*}[tb]
    \centering
    \includegraphics[width=.95\linewidth]{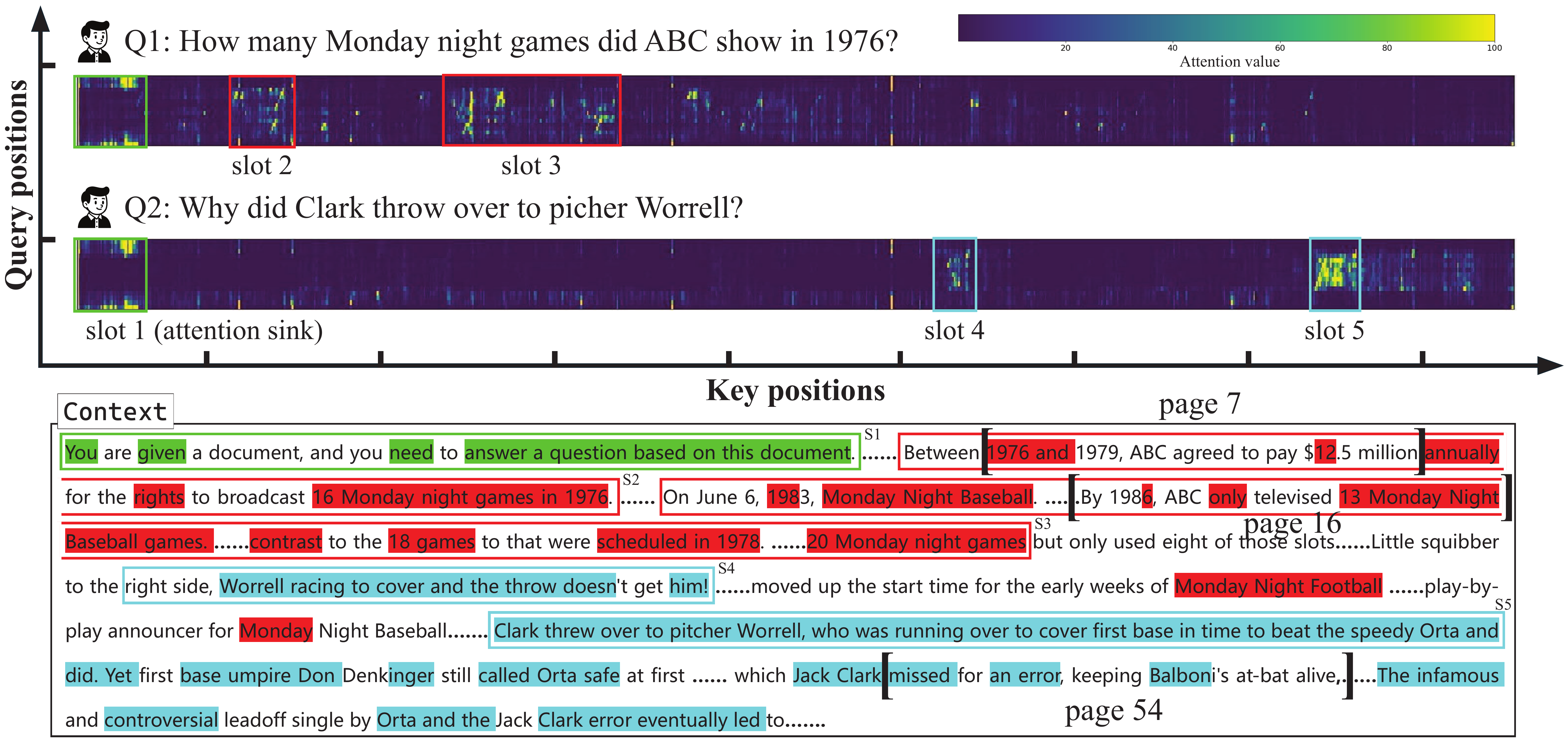}
    \vspace{-5pt}
    \caption{High-scoring tokens selected by two different queries in LLaMA \cite{grattafiori2024llama} are mapped to their corresponding text. Important tokens are query-dependent and sparsely distributed across the context, causing pages to contain a mix of important and unimportant tokens, which leads to inaccuracy in page-level retrieval.}
    \label{ob1}
    \vspace{-17pt}
\end{figure*}
\subsection{Preliminaries}
% Brief intro of KV cache. And token selection
% page-based selection algorithm
% use euqation represent the correlation between page size and selection overhead trade-off.
In an autoregressive LLM, the inference process typically consists of two stages: the prefill stage and the decoding stage.

 \noindent \textbf{Prefill Stage.} Given an input context $\mathbf{X} \in \mathbb{R}^{l_{prompt} \times d}$, the model computes the initial key and value representations, denoted as $\mathbf{K}_0 \in \mathbb{R}^{l_{prompt} \times d}$ and $\mathbf{V}_0 \in \mathbb{R}^{l_{prompt} \times d}$, which together form the initial KV cache.
 
\noindent\textbf{Decoding Stage.} At each step, a new query $\mathbf{q}\in \mathbb{R}^{1\times d }$ is generated and the corresponding key $\mathbf{k}$ and value $\mathbf{v}$ are appended to the initial cache $(\mathbf{K}_0, \mathbf{V}_0)$ to form the current KV cache:
\begin{align*}
    \mathbf{K} \xleftarrow{} \text{Concat}(\mathbf{K}_0,\mathbf{k}), \mathbf{V} \xleftarrow{} \text{Concat}(\mathbf{V}_0,\mathbf{v}).
\end{align*}
The attention output is then computed as:
\begin{align*}
    \mathbf{s} = \text{softmax}(\mathbf{q}\mathbf{K}^{T}), \quad \mathbf{o} =\mathbf{s} \mathbf{V}.
\end{align*}
The major overhead of decoding comes from computation of attention score $\mathbf{s}$, which reflects the importance of KV token $\mathbf{k}_j$ to the query. At each step, the current query must attend to the entire KV cache. This cost becomes higher in long-context tasks. To address this, previous works have demonstrated attention sparsity, observing that initial tokens (attention sink) and recent tokens (locality) tend to receive higher attention scores.
Based on this, they retain a fixed subset of tokens, denoted as $(\mathbf{K^{\prime}},\mathbf{V^{\prime}}) \in \mathbb{R}^{n\times d}$ at these positions for all queries, where $n$ is cache budget.

However, subsequent studies show that token criticality varies across different queries. As a result, query-specific token selection is necessary, where token importance needs to be recomputed for each query. To improve importance estimation efficiency, existing works tend to perform selection at a coarser, page-level granularity. For example, Quest \cite{tang2406quest} partitions the key cache $\mathbf{K} \in \mathbb{R}^{l \times d}$ into $\frac{l}{L}$ pages ($L$ is typically 16 or 32). For each page $\mathbf{P}$, it extracts maximum and minimum vectors $\mathbf{k}_P^{\max}$ and $\mathbf{k}_P^{\min} \in \mathbb{R}^{1 \times d}$, performs point-wise multiplication with $q$, 
\begin{align}
    \mathbf{\alpha}^{\max}  &= \mathbf{q} \odot \mathbf{k}_P^{\max}  ,\\
    \mathbf{\alpha}^{\min}  &= \mathbf{q} \odot \mathbf{k}_P^{\min}  , 
\end{align}
and takes the maximum across both the hidden dimension and the two vectors to obtain the page importance score.
\begin{align}
s_P = \max_{i=1,\dots,d} \left( \max \left(\mathbf{\alpha}^{\max}_i,\; \mathbf{\alpha}^{\min}_i \right) \right).
\end{align}
Pages with the highest importance scores are then selected for self-attention computation. 

During the selection, Quest loads only 2 $\mathbf{K}$ vectors per page.
Assuming $\mathbf{K}$ is stored in \texttt{float16}, this results in a key cache load ratio of:
\begin{align}
\frac{2\times16\times l/L}{l \times 16} =\frac{2}{L}.
\end{align}
It is clear that larger page sizes reduce importance estimation costs, but lead to coarser granularity by grouping more tokens together. There exists a trade-off between efficiency and accuracy.

\subsection{Fine-grained and Efficient KV Retrieval}
To improve upon existing page-level KV retrieval methods, we make the following two observations on the token importance estimation process. 

\noindent\textbf{OB1: Important Token Sparsity Makes Page Retrieval Inaccurate.} To understand the trade-off of page granularity, we visualize both the highest-attended tokens and those selected under page-level partitioning by mapping them back to the original context. As shown in Fig.~\ref{ob1}, queries $Q1$ and $Q2$ attend to different regions of the context, which aligns with prior findings on the dynamic nature of token criticality.
Moreover, tokens with high attention scores are sparsely distributed across the context, and we observe that pages 7, 16, and 54 each contain a mixture of both important and unimportant tokens.
This overlap makes inaccurate retrieval, which pinpoints the necessity of fine-grained retrieval strategies that identify important information at the token level, rather than relying on coarse-grained page grouping.

Motivated by this, we aim to design a retrieval strategy that operates at fine-grained token level while incurring minimal additional computation overhead.
Notably, quantization enables low-bit computation, achieving high efficiency while still allowing every token to participate in the criticality estimation. We begin by validating this insight through the following observation.

\noindent \textbf{OB2: Quantization has Minimal Impact on Top-$k$ Selection, even at 1-bit.}
Quantizing KV cache values to a lower precision significantly reduces the cost of data movement and attention computation. 
Let $\mathbf{K} \in \mathbb{R}^{l \times d}$ denote the original key cache, where $\mathbf{k}_i \in \mathbb{R}^{1 \times d}$ is the key vector corresponding to the $i$-th token, quantization converts each $\mathbf{k}_i$ to its dequantized counterpart $\tilde{\mathbf{k}}_i$ as
\begin{align}
\mathbf{k}_i^Q = \left\lfloor \frac{\mathbf{k}_i - \mathbf{z}_i^K}{\mathbf{s}_i^K} \right\rfloor, \quad
\tilde{\mathbf{k}}_i = \mathbf{k}_i^Q \odot \mathbf{s}_i^K + \mathbf{z}_i^K,
\end{align}
where  $\mathbf{s}_i^K, \mathbf{z}_i^K \in \mathbb{R}^{1 \times d} $ are the per-channel calibrated scaling and bias vectors specific to the $i$-th key vector.
%I think some calibration based quantization method use this objective
\begin{figure}[htbp]
     \centering
\includegraphics[width=1\linewidth]{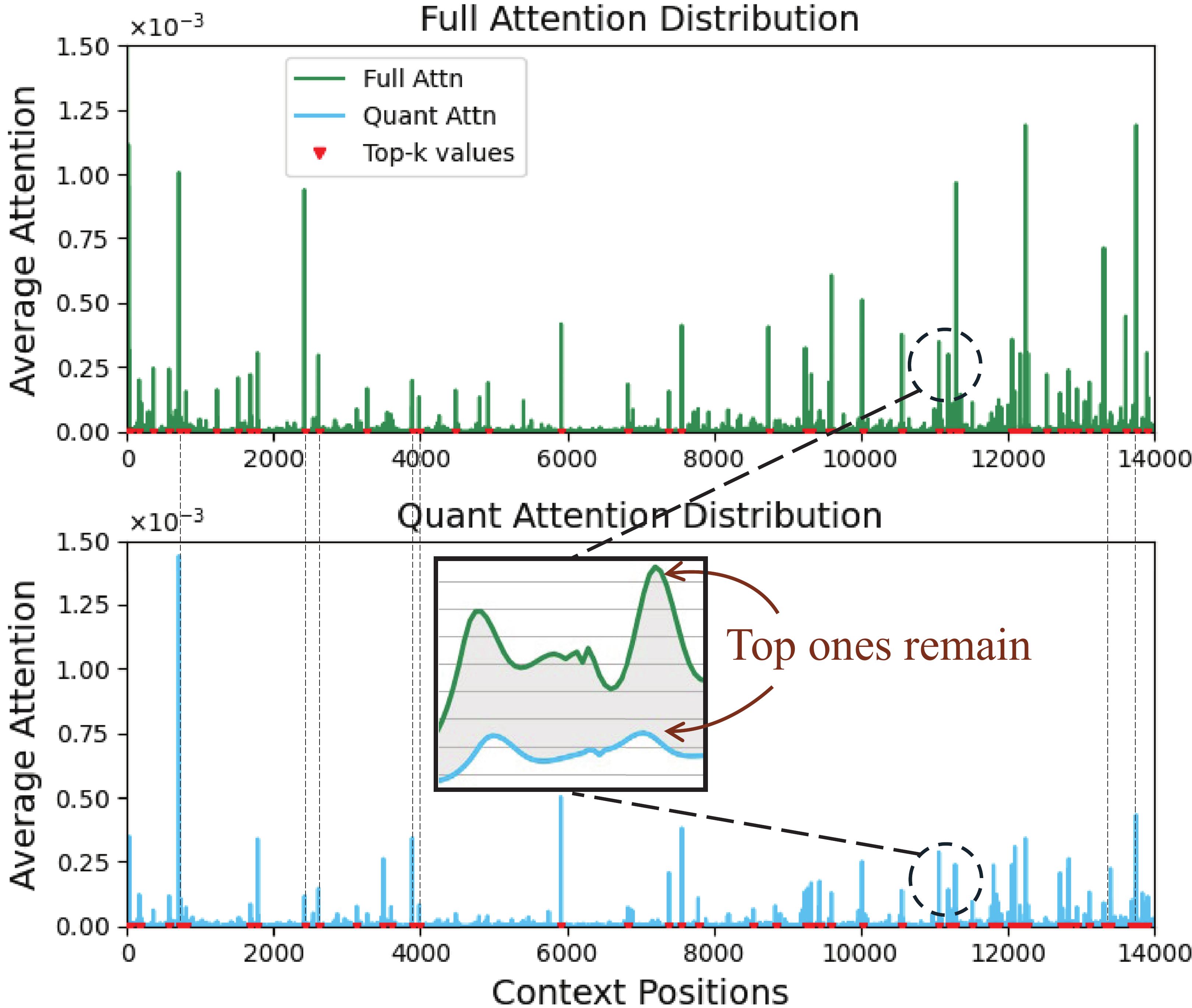}
\vspace{-20pt}
    \caption{Averaged full/quantized attention scores along the sequence. Despite distribution distortion caused by low-bit quantization, the Top-$k$ tokens are largely preserved, indicating that token criticality remains identifiable under extreme quantization.}
    \label{ob2}
    \vspace{-7pt}
\end{figure}
Previous KV quantization methods \cite{zhang2024kv} aim to optimize these factors through the calibration process to minimize the impact of quantization on the computed attention score. 
This objective can be formulated as an $\ell_2$ loss:
\begin{align}
\min_{\mathbf{s}^K, \mathbf{z}^K} \sum_{i=1}^{l} \left( \mathbf{q} \mathbf{k}_i^\top - \mathbf{q} \tilde{\mathbf{k}}_i^\top \right)^2.
\end{align}
However, quantization introduces perturbations on the values, drawing computation results away from intended. The impact of quantization is more severe if outliers exist in the distribution. Previous methods like Kivi~\cite{liu2024kivi}, equipped with advanced channel-wise grouping and rescaling scheme, cannot quantize the KV cache below 2 bit while retaining model performance. 

Meanwhile, we observe that quantizing the key cache to low bits has significantly less impact on the token importance estimation than retaining the attention scores.
Here we explore an extreme case by quantizing $\mathbf{K}$ to just 1-bit.
Using the full-precision attention scores as ground truth, we evaluate whether Top-$k$ token selection can still be accurately recovered under such an aggressive quantization setting. Specifically, we feed a long input context (14k tokens) into LlaMA during the prefill stage, and compute attention scores under both full-precision and 1-bit quantized $\mathbf{K}$ using the same query.
As shown in Fig.~\ref{ob2}, despite the fact that low-bit quantization truncates large values and distorts the overall distribution, the Top-$k$ tokens remain largely unchanged.
This suggests that token criticality is still well captured, even under extreme quantization.

\begin{figure}[h]
    \centering
    \includegraphics[width=1\linewidth]{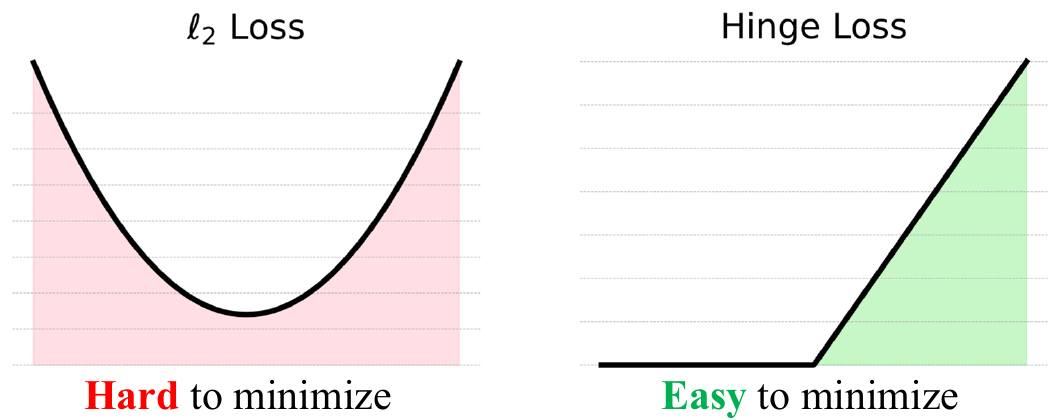}
    \vspace{-20pt}
    \caption{Intuition of Fier. Our relaxed quantization objective ignores errors smaller than $m/2$, allowing the use of extremely low bit-widths while preserving Top-$k$ ranking accuracy. }
    \label{hinge}
    \vspace{-19pt}
\end{figure}

To understand the reason, we analyze the quantization objective implied by the goal of token importance estimation. 
For importance estimation, we aim to maintain the ranking of the Top-$k$ tokens rather than preserving all attention scores precisely. 
Assume $m$ is the minimum margin between the attention scores of Top-$k$ and non-Top-$k$ tokens in the full-precision setting.
To preserve this ranking after quantization, it is sufficient to ensure that the attention score of each token deviates from its full-precision counterpart by at most $m/2$.
This leads to the following hinge objective:
\begin{align}
\small
\min_{\mathbf{s}^K,\mathbf{z}^K} \sum_{i=1}^l  \max\left(0,\; \frac{m}{2} - \left(\mathbf{q} {\mathbf{k}}_i^\top - \mathbf{q} \tilde{\mathbf{k}}_i^\top\right) \right).
\end{align}
Compared to the $\ell_2$ loss, the hinge loss imposes a relaxed objective that prioritizes preserving the relative ranking of Top-$k$ tokens (Fig.~\ref{hinge}). More importantly, outlier tokens that lead to large attention scores enjoy larger margins under the hinge loss, making their quantization errors less impactful.
This enables quantization with 1-bit and larger group sizes $g$ while still maintaining Top-$k$ accuracy. 
\subsection{Fier Workflow}
Motivated by previous observations, we propose Fier, a token-level retrieval method based on 1-bit linear quantization.
Fier compresses the key cache into 1-bit using a simple round-to-nearest (RTN) quantizer.
Given a query $\mathbf{q}$, approximate attention scores are computed efficiently using quantized keys. Based on these scores, Fier selects the Top-$k$ tokens and performs full-precision self-attention over the selected subset.
We summarize the workflow of Fier in Algorithm \ref{alg1}.

\subsection{Theoretical Analysis of Efficiency}
Beyond retrieval accuracy, the efficiency of the selection and decoding stage is also critical for practical deployment. We measure the efficiency using the key cache access ratio (CAR), defined as the fraction of the key cache accessed throughout the pipeline. Specifically, we compare both retrieval- and eviction-based methods under the same top-$k$ setting. In the decoding phase, all methods (Fier, Quest, and H2O) access $k$ tokens, so their CAR is identical.
The key difference lies in the selection phase.
For $\mathbf{K}$ stored in \texttt{float16}, we quantize it to 1-bit with group size $g$. Note that in addition to the 1-bit $\mathbf{K}_Q$, each group also needs to store a pair of $(s, z)$ in \texttt{float16}. Therefore, the CAR of Fier will be calculated as:
\begin{align}
    \frac{l\times 1+(l/g)\times2\times16}{l\times16}=\frac{1+32/g}{16},
\end{align}
which decreases with a larger group size. 
Recall that Quest has a load ratio of $2/L$. For fairness, we set $g = 32$, which matches the load ratio of $1/8$ with page size $L = 16$ as implemented in the Quest baseline. 
In contrast, H2O maintains a dynamic pool of size $k/l$ during the selection stage. Therefore, when $k = l/10$, the CAR of the three methods can be summarized in Tab.~\ref{tab:car}. While retrieval does introduce additional overhead during the selection compared to eviction, its goal is to improve performance with only a modest CAR increase.
\begin{algorithm}[!t]
\normalsize
\caption{Fier: Token-Level KV Retrieval via 1-Bit RTN Quantization}
\label{alg1}
\begin{algorithmic}[1]
\small
\State \textbf{Input:} Query $\mathbf{q}$, full-precision $(\mathbf{K},\mathbf{V})$, group size $g$
\State \textbf{Output:} Attention output $\mathbf{o}$

\State \textcolor{green!50!black}{// Step 1: Quantize $\mathbf{K}$ to 1-bit}
\State Partition $\mathbf{K}$ into groups of size $g$ along each channel
\State For each group, compute the scaling factors $(s, z)$ and broadcast them to construct \( \mathbf{s}^K, \mathbf{z}^K \in \mathbb{R}^{l \times d} \).
\State $\mathbf{K}_Q = \left\lfloor \frac{\mathbf{K} - \mathbf{z}^K}{\mathbf{s}^K} \right\rfloor, \mathbf{K}_Q \in \{-1, 1\}^{l \times d}
$ \textcolor{green!50!black}{\# binary}

\State $\tilde{\mathbf{K}} = \mathbf{K}_Q \odot \mathbf{s}^K + \mathbf{z}^K$

\State \textcolor{green!50!black}{// Step 2: Compute Approximate Attention Scores}
\State $\tilde{\mathbf{s}} = \mathbf{q} \cdot \tilde{\mathbf{K}}^\top$

\State \textcolor{green!50!black}{// Step 3: Select Top-$k$ Tokens}
\State $\mathcal{S}_q = \text{Top-}k(\tilde{\mathbf{s}})$

\State \textcolor{green!50!black}{// Step 4: Compute Real Attention on Selected Tokens}
\State $\mathbf{K}' = \mathbf{K}[\mathcal{S}_q], \mathbf{V}' = \mathbf{V}[\mathcal{S}_q]$
\State \textbf{Return:} $\mathbf{o} = \text{softmax}(\mathbf{q} \mathbf{K}'^\top)  \mathbf{V}'$
\end{algorithmic}

\end{algorithm}
\vspace{-6pt}

\begin{table}[b]
\vspace{-6pt}
\footnotesize
\centering
\caption{CAR comparison of different methods (Assuming $k = l/10$).}
\vspace{-5pt}
\begin{tabular}{cccc}
\toprule
Method & Sel. (\%) & Dec. (\%) & Total CAR (\%) \\
\midrule
H2O   & 10.0 & 10.0 & 20.0 \\
Quest & 12.5 & 10.0 & 22.5 \\
Fier  & 12.5 & 10.0 & 22.5 \\
\bottomrule
\end{tabular}
\label{tab:car}
\vspace{-8pt}
\end{table}

%Under this setting, for a context of length 32k and a budget of $n=\text{2k}$, Fier selects the Top-2k KV tokens for real self-attention computation. The total KV cache load ratio is:
%\begin{align}
%   \frac{1}{2\times8 } + \frac{2\times 2\text{k}}{2\times32\text{k}} = \frac{1}{8},
%\end{align}
%which indicates that Fier can reduce memory load by 8$\times$ compared to using the full KV cache.
\begin{figure*}[t]
     \centering
    \includegraphics[width=.9\linewidth]{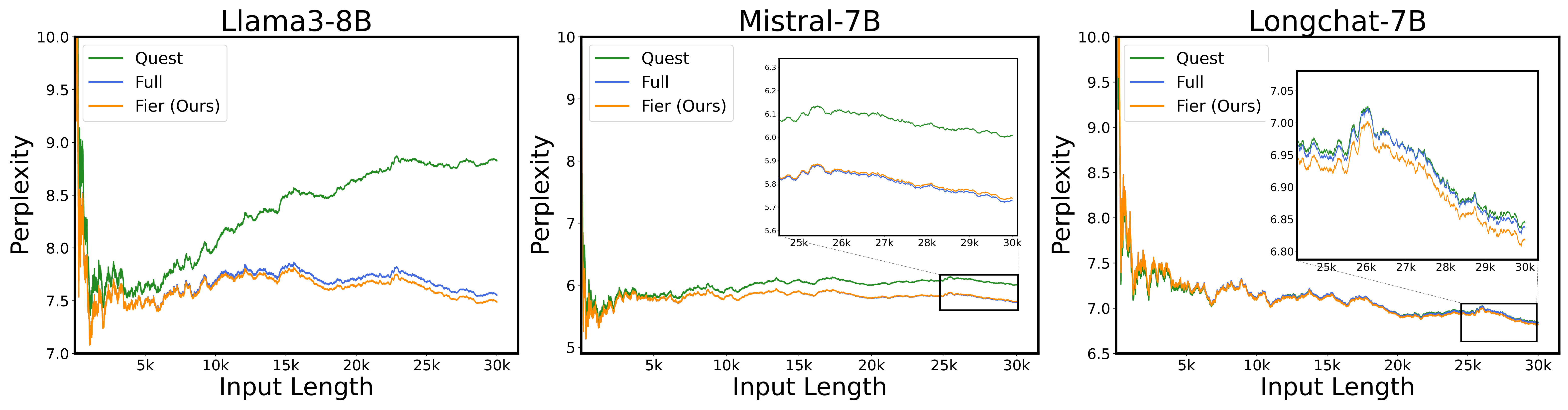}
    \vspace{-9pt}
    \caption{Language modeling evaluation. We measure output perplexity by prompting the model with input lengths ranging from 0 to 32k tokens. Fier achieves performance comparable to full KV and significantly surpasses Quest. }
    \label{exp1}
    \vspace{-16pt}
\end{figure*}
\section{Experiments}
\subsection{Setting}
\noindent \textbf{Datasets.}
We evaluate Fier on the language modeling task PG19 \cite{rae2019compressive}. To assess its performance on long-context QA, we further conduct experiments on LongBench \cite{bai2023longbench} using six representative datasets: NarrativeQA, HotpotQA, Qasper, TriviaQA, GovReport, and MultifieldQA. The detailed information about the six datasets
is in Appendix \ref{AppendixA}. We evaluate Fier on the passkey retrieval task \cite{peng2023yarn} to assess its ability to model long-range dependencies. We also compare responses from Fier and Quest enabled chatbots in Appendix~\ref{AppendixC}.

\noindent \textbf{Models.} We apply our method to three open-sourced models: LLaMA-3-8B-Instruct \cite{grattafiori2024llama}, LongChat-v1.5-7B-32k \cite{li2023long}, and Mistral-7B-Instruct \cite{jiang2023mistral}. Following the same setup as in Quest, neither Fier nor the baseline methods are applied to the first two layers of the model. We evaluate the performance of each method under varying KV cache budgets.

\noindent \textbf{Baselines.} 
We thoroughly compare the performance of Fier and Quest \cite{tang2406quest} across various benchmarks. Note that in all performance evaluation, we set the page size of Quest to 16 and the grouping size of Fier to 32 for a fair comparison.
We also compare Fier with four KV eviction baselines: H2O \cite{zhang2023h2o}, StreamingLLM \cite{xiao2023efficient}, SnapKV \cite{li2024snapkv} and TOVA \cite{oren2024transformers}. The results in the experiments are either taken from the original paper or obtained by running open-source code. More implementation details can be found in Appendix \ref{AppendixB}.

% IN 3 
\begin{figure}[htbp]
     \centering
    \includegraphics[width=1\linewidth]{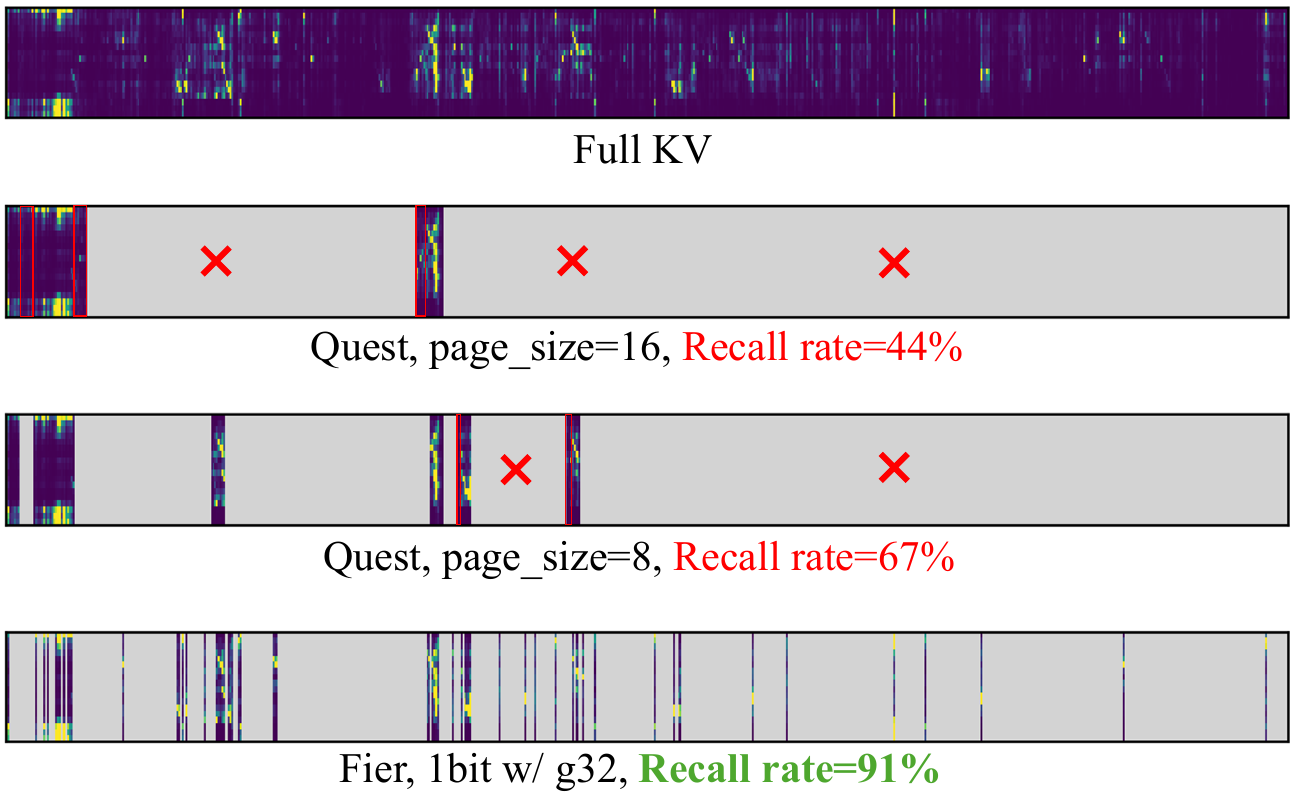}
    \vspace{-20pt}
    \caption{Fier’s token-level retrieval preserves more Top-$k$ tokens compared to Quest’s page-level approach, resulting in higher recall and better alignment with full attention.}
    \label{insgiht}
    \vspace{-20pt}
\end{figure}
\subsection{Insight Verification}
\textbf{More Accurate Retrieval of Important Tokens.} In Fig.~\ref{insgiht}, we visualize the positions of Top-$64$ tokens selected by Quest with different page sizes and by Fier-1bit-g32, all mapped back onto the full KV cache. We then compute the recall rate, defined as the overlap between the retrieved tokens and those selected using the full attention score. The experiment is conducted on LLaMA.

We observe that Quest with either small or large page sizes tends to retain unimportant tokens and evict important ones, due to its coarse-grained page-level retrieval. In contrast, Fier performs token-level retrieval through low-bit quantized attention computation, resulting in a significantly higher recall rate and better alignment with the full attention distribution.
\begin{figure*}[tb]
     \centering
    \includegraphics[width=.9\linewidth]{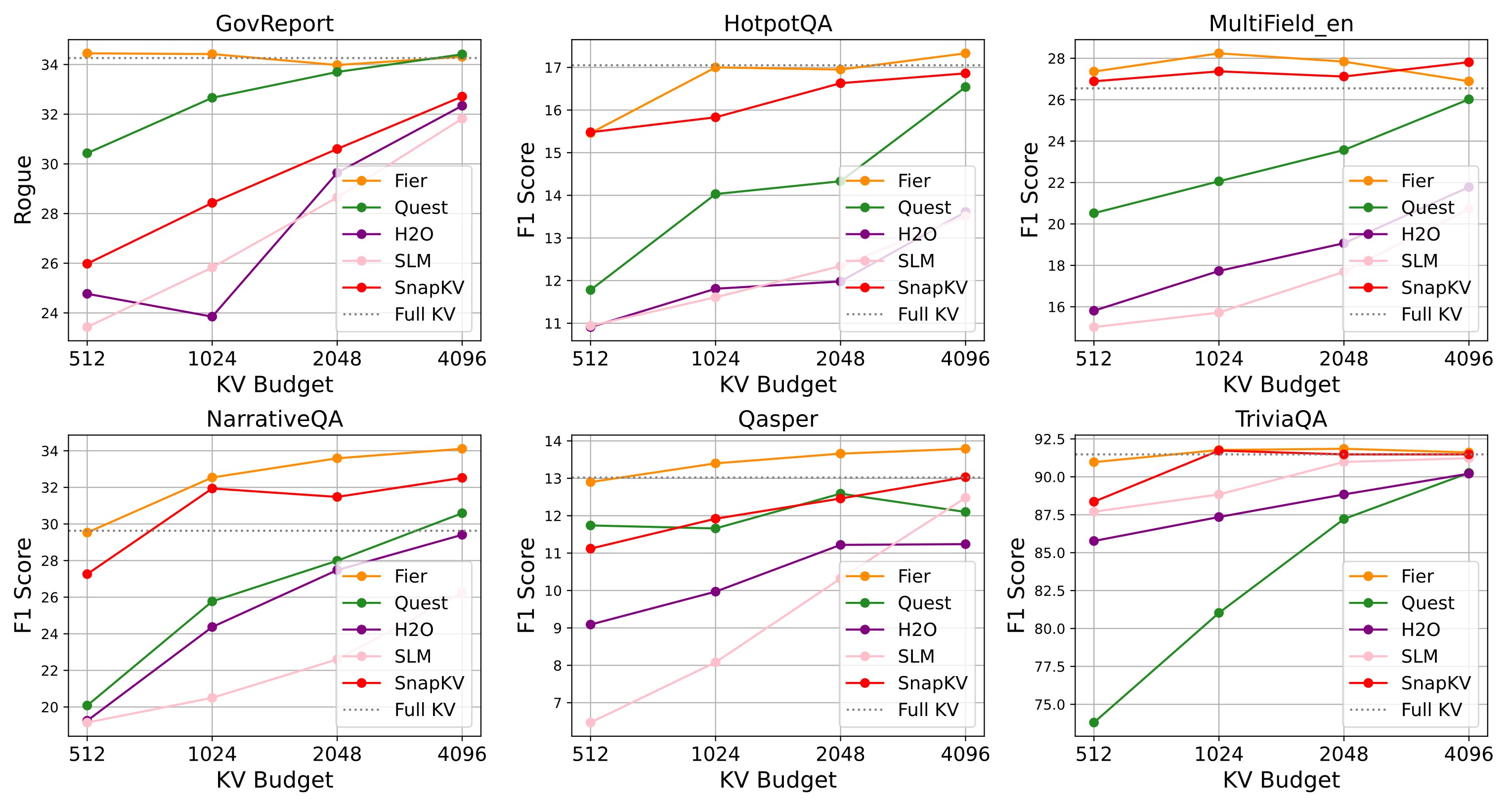}
        \vspace{-5pt}
    \caption{LongBench evaluation on LaMA-3-8B-Instruct. Fier outperforms all baselines across six long-context datasets and matches full KV performance with just 1k cache budget. }
    \label{exp2}
    \vspace{-7pt}
\end{figure*}
\subsection{Performance Evaluation}

\begin{table*}[h]
\scriptsize
\centering
\captionsetup{justification=centering, singlelinecheck=false}
\caption{LongBench evaluation on LongChat and Mistral. Consistent performance gains of Fier on two models.}
\vspace{-5pt}
\begin{tabular}{ccccccccccccccc}
\toprule
\multirow{2}{*}{\textbf{LLMs}} & \multirow{2}{*}{\textbf{Method}}& \multicolumn{4}{c}{\textbf{Multifield\_en}} 
& \multicolumn{4}{c}{\textbf{NarrativeQA}} 
& \multicolumn{4}{c}{\textbf{GovReport}} & \multirow{2}{*}{\textbf{Avg.}}\\
\cmidrule(lr){3-6} \cmidrule(lr){7-10} \cmidrule(lr){11-14}
& & 512 & 1024 & 2048 & 4096 & 512 & 1024 & 2048 & 4096 & 512 & 1024 & 2048 & 4096 \\
\midrule
 \multirow{6}{*}{LongChat-7B} &
\cellcolor{gray!10}Full KV  & \multicolumn{4}{c}{\cellcolor{gray!10}43.2}
  & \multicolumn{4}{c}{\cellcolor{gray!10}20.88}
  & \multicolumn{4}{c}{\cellcolor{gray!10}30.89} & \cellcolor{gray!10}31.66 \\
& SLM  & 21.17 & 21.29 & 26.55 & 34.82 & 10.69 & 12.46 & 17.55 & 18.94 & 16.85 & 21.88 & 22.77 & 26.96 &20.99\\
& H2O  & 21.15 & 25.07 & 30.28 & 37.75 & 10.67 & 12.96 & 14.75 & 19.31 & 19.73 & 22.69 & 26.15 & 27.55 &22.34\\
& SnapKV & 36.74 & 37.93 & 40.26 & 42.21 & \textbf{19.21} & \textbf{19.32} & 19.28 & \textbf{20.68} & 22.57 & 23.45 & 26.3 & 28.55 & 28.04\\
& Quest & 38.05 & \textbf{41.95} & \textbf{44.03} & 42.41 & 16.51 & 18.76 & 19.37 & 20.12 & 27.54 & 30.12 & \textbf{31.27} & 31.21 &30.11\\
& \cellcolor{orange!20}Fier  
  & \cellcolor{orange!20}\textbf{39.05} 
  & \cellcolor{orange!20}39.85 
  & \cellcolor{orange!20}42.4 
  & \cellcolor{orange!20}\textbf{42.54} 
  & \cellcolor{orange!20}17.23 
  & \cellcolor{orange!20}17.83 
  & \cellcolor{orange!20}\textbf{19.96} 
  & \cellcolor{orange!20}19.51 
  & \cellcolor{orange!20}\textbf{30.18} 
  & \cellcolor{orange!20}\textbf{30.89} 
  & \cellcolor{orange!20}30.85 
  & \cellcolor{orange!20}\textbf{31.67} & \cellcolor{orange!20}\textbf{30.16} \\

\midrule

\multirow{6}{*}{Mistral-7B} 
&\cellcolor{gray!10}Full KV & \multicolumn{4}{c}{\cellcolor{gray!10}52.92} & \multicolumn{4}{c}{\cellcolor{gray!10}28.49} & \multicolumn{4}{c}{\cellcolor{gray!10}34.81} &\cellcolor{gray!10}38.74 \\
& SLM & 29.91 & 31.16 & 35.75 & 44.12 & 24.21 & 24.79 & 25.91 & 28.9 & 22.09 & 24.6 & 27.57 & 31.19 &29.18\\
& H2O & 47.39 & 48.43 & 49.03 & 49.95 & 23.04 & 27.79 & 28.6 & 30.2 & 24.24 & 26.15 & 27.19 & 30.04 &34.34\\
& SnapKV & 53.05 & 52.64 & 52.92 & \textbf{53.44} & 25.57 & 28.09 & \textbf{30.27} & 29.76 & 25.83 & 28.28 & 30.91 & 32.74 &36.96\\
& Quest & 48.07 & 50.67 & 53.7 & 51.76 & 20.25 & 25.71 & 28.31 & 27.28 & 31.42 & 32.57 & 33.07 & 33.52 &36.36\\
& \cellcolor{orange!20}Fier  
  & \cellcolor{orange!20}\textbf{53.97} 
  & \cellcolor{orange!20}\textbf{54.67} 
  & \cellcolor{orange!20}\textbf{54.37} 
  & \cellcolor{orange!20}53.32 
  & \cellcolor{orange!20}\textbf{26.75 }
  & \cellcolor{orange!20}\textbf{28.75} 
  & \cellcolor{orange!20}29.11 
  & \cellcolor{orange!20}\textbf{31.11} 
  & \cellcolor{orange!20}\textbf{34.47} 
  & \cellcolor{orange!20}\textbf{34.53} 
  & \cellcolor{orange!20}\textbf{34.65} 
  & \cellcolor{orange!20}\textbf{34.9} & \cellcolor{orange!20}\textbf{39.22}\\
\bottomrule
\end{tabular}
\label{tab:exp2}
\vspace{-12.55pt}
\end{table*}

\subsubsection{PG19 Results}
We begin by evaluating language modeling perplexity on PG19, a benchmark consisting of 100 books with an average length of 70k tokens. We evaluate three different models by feeding them texts of varying lengths and compare the results against both Full KV and Quest (Fig.~\ref{exp1}). Note that both Fier and Quest are evaluated under the same KV cache budget of 4096 tokens. Fier achieves performance close to that of Full KV and significantly outperforms Quest on both the LLaMA and Mistral models.
\subsubsection{Longbench Results}
We evaluate on the LongBench benchmark using LLaMA-3-8B-Instruct across a diverse set of long-context tasks, including single-document QA (NarrativeQA, Qasper, MultiFieldQA), multi-document QA (HotpotQA), summarization (GovReport), and few-shot learning (TriviaQA). We also compare Fier with H2O, StreamingLLM, SnapKV, and Quest under varying KV cache budget settings. In addition, we perform evaluations using the Mistral-7B and LongChat-7B models to verify the generality of our method across different model architectures.

As shown in Fig.~\ref{exp2}, Fier consistently achieves superior performance compared to all baselines across six long-context datasets under various KV cache budgets. Overall, Fier surpasses all baselines at cache budgets of 512, 1024, and 2048 tokens, and achieves comparable performance to full KV using only 1k tokens, which is only 11\% of the full cache. This suggests that Fier benefits from accurately retrieving critical tokens, enabling efficient use of limited cache resources without compromising model quality. Similar results are observed on the other two models. As shown in Tab.~\ref{tab:exp2}, Fier shows consistent improvements on both single-document and multi-document tasks. 
% Examples of response quality can be found in Appendix \ref{AppendixC}.

\begin{figure*}[h]
     \centering
    \includegraphics[width=.9\linewidth]{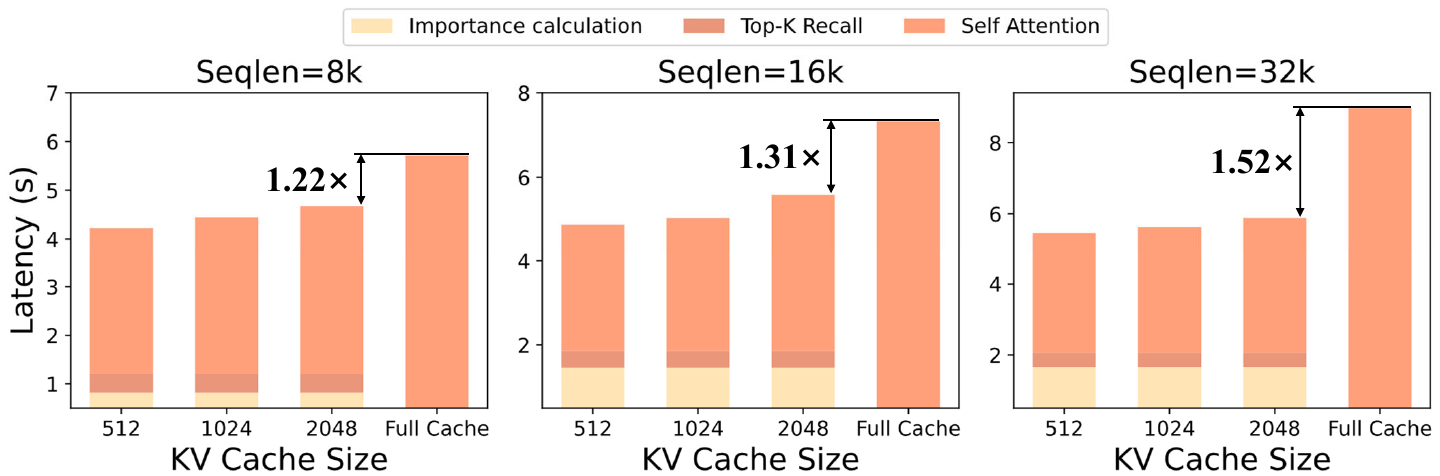}
    %\vspace{-10pt}
    \caption{Decoding latency of 256 tokens on LLaMA-2-7B \cite{touvron2023llama} under varying prefill context lengths. Fier achieves increasing speedup over full KV by restricting attention to a small subset of the cache. At 32k context length, it delivers over 1.5× acceleration.}
    \label{exp4}
    \vspace{-10pt}
\end{figure*}
\subsubsection{Passkey Retrieval}

We further evaluate Fier’s ability to handle long-distance dependencies. Specifically, we employ the passkey retrieval task \cite{peng2023yarn} as our benchmark. This task assesses whether the model can retrieve a simple passkey from a large amount of irrelevant content. Following the setup in \citet{tang2406quest}, we use a context length of 10k tokens for evaluation with both LongChat-7B and a smaller model, LLaMA3-1B. KV eviction methods perform poorly due to their inability to recall discarded tokens, while Quest provides noticeable improvements over them. Fier, however, achieves even higher retrieval accuracy, performing well under extremely low budgets of just 32 and 64 tokens, especially improving the long-context processing capability of smaller models (shown in Tab.~\ref{exp3}).
\begin{table}[!t]
    \centering
    \captionsetup{justification=justified} 
    \caption{Passkey retrieval accuracy under 10k context length. KV eviction methods struggle to recall discarded tokens, while Quest improves retrieval performance. Fier achieves the highest accuracy, even with extremely low budgets (32 and 64), effectively enhancing smaller models.}
    \label{exp3}
    \vspace{-4pt}
\small
    \begin{tabular}{cccccccc}
    \toprule
    \textbf{Longchat-7B}  & \multicolumn{5}{c}{\textbf{Cache Budget}} \\
    \textbf{Method} & 32 & 64 & 128 & 256 &512\\
    \midrule
    H20 & 0\% & 1\% & 1\% &1\% &3\%\\
    StreamingLLM & 1\% & 1\% & 1\% &3\% &5\%\\
    TOVA & 0\% & 1\% & 1\% &3\% &8\%\\
    Quest& 65\% & 99\% & 99\% &99\% &100\%\\
     \cellcolor{orange!20}Fier (ours)&  \cellcolor{orange!20}\textbf{87}\% &  \cellcolor{orange!20}\textbf{99}\% &  \cellcolor{orange!20}\textbf{99}\% &  \cellcolor{orange!20}\textbf{100}\% & \cellcolor{orange!20}\textbf{100}\%\\
\bottomrule
\end{tabular}
 \begin{tabular}{cccccccc}
    \toprule
    \textbf{LlaMA3-1B}  & \multicolumn{5}{c}{\textbf{Cache Budget}} \\
    \textbf{Method} & 32 & 64 & 128 & 256 &512\\
    \midrule
    H20 & 0\% & 1\% & 0\% &1\% &2\%\\
    StreamingLLM & 0\% & 0\% & 1\% &2\% &4\%\\
    TOVA & 0\% & 1\% & 1\% &3\% &6\%\\
    Quest& 36\% & 60\% & 92\% &99\% &99\%\\
    \cellcolor{orange!20}Fier (ours)& \cellcolor{orange!20}\textbf{63}\% & \cellcolor{orange!20}\textbf{87}\% & \cellcolor{orange!20}\textbf{97}\% & \cellcolor{orange!20}\textbf{100}\% &\cellcolor{orange!20}\textbf{100}\%\\
\bottomrule
\end{tabular}
\vspace{-14pt}
\end{table}

\subsection{Ablation Study}
\textbf{Token Granularity vs. Quantized Attention.} To understand whether Fier’s performance gain primarily stems from its fine-grained token-level selection (as opposed to page-level) or from the use of quantized attention as an importance metric, we conduct an ablation study on LLaMA-3-8B-Instruct, isolating these two factors. Specifically, we reduce the page size in Quest to approximate finer granularity and compare performance. In addition, we apply the averaged quantized attention score as the page-level importance metric under the same page size, and evaluate its effect on Quest.
As shown in Tab.~\ref{tab:ablation}, using smaller page sizes in Quest leads to improved performance. However, it also increases the cache load ratio. Additionally, incorporating quantized attention for scoring further enhances its effectiveness. Notably, Fier can be viewed as quantized attention with a page size of 1, achieving the best overall results. These results suggest that Fier benefits from both the token-level granularity and the use of quantized attention as a lightweight yet effective importance estimator.

\noindent\textbf{Fier w/ Different Group Sizes.} We also investigate how the group size used during key quantization affects Fier’s performance. We find that as the group size increases, the cache load ratio decreases, but this comes at the cost of reduced performance. Nevertheless, Fier consistently outperforms Quest under the same cache load ratio.

\begin{table}[!t]
\footnotesize
\centering
\captionsetup{justification=justified} 
\caption{Ablation study. Fier benefits from both token granularity and quantized attention. Larger group sizes yield better efficiency but may reduce accuracy.}
\vspace{-2pt}
\begin{tabular}{ccccc}
\toprule
  \multirow{2}{*}{\textbf{Method}} &\multirow{2}{*}{\textbf{Load R.}}& \multicolumn{3}{c}{\textbf{HotpotQA}} \\
\cmidrule(lr){3-5}
 & & 512 & 1024 & 2048 \\
\midrule
 Quest-p32 & 0.063 & 7.16 &8.98 & 11.39   \\
 Quest-p16 & 0.125 & 11.78 & 14.03 & 14.33   \\
Quest-p16-w/quant & 0.125 & 13.54 & 14.77 & 15.26   \\
 Quest-p8 &0.25 & 15.16 & 16.66 & \textbf{17.3}  \\
 % Quest-p8-w/quant &1/8 & null & null & null & null\\
 \midrule
 % Quest-p4 &1/4 & 17.43 & 17.42 & 17.44 & 17.18  \\
 Fier-g256 & 0.07 & 12.55 &  14.51 &16.73   \\
 Fier-g128 & 0.08 & 13.96 & 15.53 & 17.04   \\
 \cellcolor{orange!20}Fier-g32  & \cellcolor{orange!20}0.125
  & \cellcolor{orange!20}\textbf{15.46} 
  & \cellcolor{orange!20}\textbf{17.0} 
  & \cellcolor{orange!20}16.95
   \\
\bottomrule
\end{tabular}
\label{tab:ablation}
\vspace{-15pt}
\end{table}
\subsection{Efficiency Profiling}
\textbf{Inference Efficiency.} In Fig.~\ref{exp4}, we present the decoding latency of 256 tokens on LLaMA-2-7B under different prefill context lengths. To ensure a fair comparison with Full KV, we include both the time spent on computing quantized attention and the time required to recall the selected Top-$k$ tokens. We implement the group-wise quantization kernel using Triton, and employ the Top-$k$ CUDA operator to efficiently perform Top-$k$ token recall. Fier’s efficiency gain is mainly attributed to the speedup in the self-attention computation, as it restricts attention to only a small subset of the KV cache. This acceleration becomes more pronounced as the context length increases; for instance, at a context length of 32k tokens, Fier achieves over 1.5× decoding speedup. We also provide a detailed latency breakdown comparison of Fier and Quest in Appendix \ref{AppendixE}.

\section{Conclusion}
We present Fier, a fine-grained and efficient KV cache retrieval algorithm that selects important tokens using 1-bit quantization. By involving all tokens in the computation, Fier enables token-level criticality estimation, leading to improved recall rate and model performance. Extensive experiments across various tasks and model architectures show that Fier consistently outperforms existing methods. Notably, Fier matches full cache performance using only 11\% of the KV budget and achieves a 1.2$\times$–1.5$\times$ decoding speedup.
\section*{Limitations}
\noindent \textbf{Model Scale.} Due to limited computational resources, our experiments are restricted to models up to 8B parameters. Evaluating Fier on larger models (e.g., 13B, 70B) may reveal further insights into its scalability and effectiveness.

\noindent \textbf{System Optimization.} Our current implementation uses Triton to develop low-bit operators for quantized attention. While Triton offers flexibility and ease of development, it does not match the low-level optimization potential of custom CUDA kernels, potentially limiting the achievable inference speedup.

\noindent \textbf{Compatibility with GQA.} Fier is not yet integrated with grouped-query attention (GQA) \cite{ainslie2023gqa}. This is because token pruning and grouped-query attention are orthogonal in principle: GQA reduces the number of KV heads, while token pruning reduces the number of tokens. Exploring their compatibility remains an important direction for future work.
 
\section*{Acknowledgments}
This work was based upon High Performance Computing (HPC) resources supported by the University of Arizona TRIF, UITS, and Research, Innovation, and Impact (RII) and maintained by the UArizona Research Technologies department and the computational resource supported by TetraMem Inc. 
Partial support for this work was provided by NSF grants CNS-2122320, CNS-2133267, and CNS-2328972. Additional support was received through the Amazon Research Award, Cisco Faculty Award, UNC Accelerating AI Awards, NAIRR Pilot Award, OpenAI Researcher Access Award, and the Gemma Academic Program GCP Credit Award.

\bibliography{custom}

%\newpage

\appendix
\section{Dataset Details}
\label{AppendixA}
We use a subset of LongBench \cite{bai2023longbench} for long-context QA evaluation. Tab.~\ref{tab:dataset} shows the statistics and evaluation metrics used in the experiments.
\begin{table}[h]
\centering
\small
\captionsetup{justification=centering, singlelinecheck=false}
\caption{Dataset Statistics and Evaluation Metrics}
\centering
\vspace{-8pt}
\begin{tabular}{ccccc}
\toprule
\textbf{Dataset} & \textbf{Avg len} & \textbf{Metric} & \textbf{\#data} \\
\midrule
 NarrativeQA         & 18,409 & F1            & 200 \\
 Qasper              & 3,619  & F1            & 200 \\
 MultiFieldQA-en     & 4,559  & F1            & 150 \\
 HotpotQA            & 9,151  & F1            & 200 \\
 GovReport           & 8,734  & Rouge-L       & 200 \\
 TriviaQA            & 8,209  & F1            & 200 \\
\bottomrule
\end{tabular}
\vspace{-15pt}
\label{tab:dataset}
\end{table}

\section{Implementation Details}
\label{AppendixB}
For experiments on LongBench \cite{bai2023longbench} and PG19 \cite{rae2019compressive}, we use three models: LLaMA-3-8B-Instruct \cite{grattafiori2024llama}, LongChat-v1.5-7B-32k \cite{li2023long}, and Mistral-7B-Instruct \cite{jiang2023mistral}, with the maximum input length uniformly set to 32k tokens to ensure a fair comparison. All inference runs are conducted on NVIDIA A6000 GPUs. During the self-attention phase, we utilize FlashAttention \cite{dao2023flashattention} for acceleration, except for H2O \cite{zhang2023h2o}, which requires computation of historical attention scores and therefore cannot benefit from FlashAttention. 

Except for H2O \cite{zhang2023h2o} and Quest \cite{tang2406quest}, which are run using their publicly released implementations, all other baselines are run using the unified KV Cache-Factory framework \cite{cai2024pyramidkv}.
Experiments on observations and decoding latency are conducted separately on NVIDIA RTX 4090 GPUs.

 \section{Comparison of Fier and MiKV}
 \label{AppendixD}
 Different from the methods discussed in the main text, MiKV \cite{yang2024no} adopts a hybrid strategy: during decoding, it stores top tokens in high precision while retaining the remaining tokens in a low-bit format. The main distinction between Fier and MiKV is that MiKV maintains the entire KV cache in mixed precision, whereas Fier only preserves top tokens during decoding. To ensure fairness, we evaluated Fier on the MMLU dataset \cite{hendrycks2020measuring} using the same cache size as reported for MiKV. Note that the cache size of Fier accounts for the CAR in the selection phase, which is not explicitly stated in the MiKV. As shown in Tab.~\ref{tab:MiKV}, Fier demonstrates better performance compared to MiKV under the same cache budget.
\begin{table}[h]
\vspace{-6pt}
\small
\centering
\caption{Comparison of Fier and MiKV on MMLU.}
\vspace{-5pt}
\begin{tabular}{ccc}
\toprule
Cache Size (\%) & Method & Acc. (\%) \\
\midrule
25   & MiKV & 43.9 \\
\cellcolor{orange!20}{25} & \cellcolor{orange!20}Fier  & \cellcolor{orange!20}\textbf{45.12}  \\
20  & MiKV & 42.7  \\
\cellcolor{orange!20}20  & \cellcolor{orange!20}Fier & \cellcolor{orange!20}\textbf{43.75}  \\
\bottomrule
\end{tabular}
\label{tab:MiKV}
\vspace{-8pt}
\end{table}

\begin{table*}[htbp]
\centering
\small
\caption{Time breakdown of Fier and Quest with different Top-$k$.}
\begin{tabular}{cccc}
\toprule
Stage & Fier ($k=2048$) & Fier ($k=512$) & Quest ($k=2048$) \\
\midrule
Importance Estimation & 16$\mu$s & 11$\mu$s & 14$\mu$s \\
Top-$k$ Recall        & 41$\mu$s & 25$\mu$s & 23$\mu$s \\
Top-$k$ Self-Attn     & 68$\mu$s & 45$\mu$s & 68$\mu$s \\
End-to-End Latency    & 20.5ms   & \textbf{16.7ms} & 18.3ms \\
GovQA Perf.           & \textbf{34.42} & 33.98 & 33.7 \\
\bottomrule
\end{tabular}
\label{tab:latency}
\end{table*}

\section{Latency breakdown comparison with Quest}
\label{AppendixE}
We provide a detailed latency breakdown between the pipelines of Fier and Quest in Tab.~\ref{tab:latency}.
The primary difference lies in the importance estimation stage:
Fier uses an extreme 1-bit quantized attention,
while Quest adopts page-level attention.
In the top-$k$ recall stage:
Quest retrieves pages, while Fier directly retrieves top-$k$ tokens.
The final top-$k$ self-attention is identical for both methods. In Tab.~\ref{tab:latency}, we compare the average per-layer latency for each token, as well as the end-to-end generation latency. With the same $k$, Fier has higher latency than Quest due to its fine-grained top-$k$
sorting in the recall stage.
However, Fier achieves better performance. By reducing $k$
, Fier can achieve lower latency than Quest while still maintaining better performance. Overall, Fier offers a better trade-off between latency and performance compared to Quest.

\section{Comparison of Chatbot Responses}
\begin{figure*}[ht]
     \centering
    \includegraphics[width=.7\linewidth]{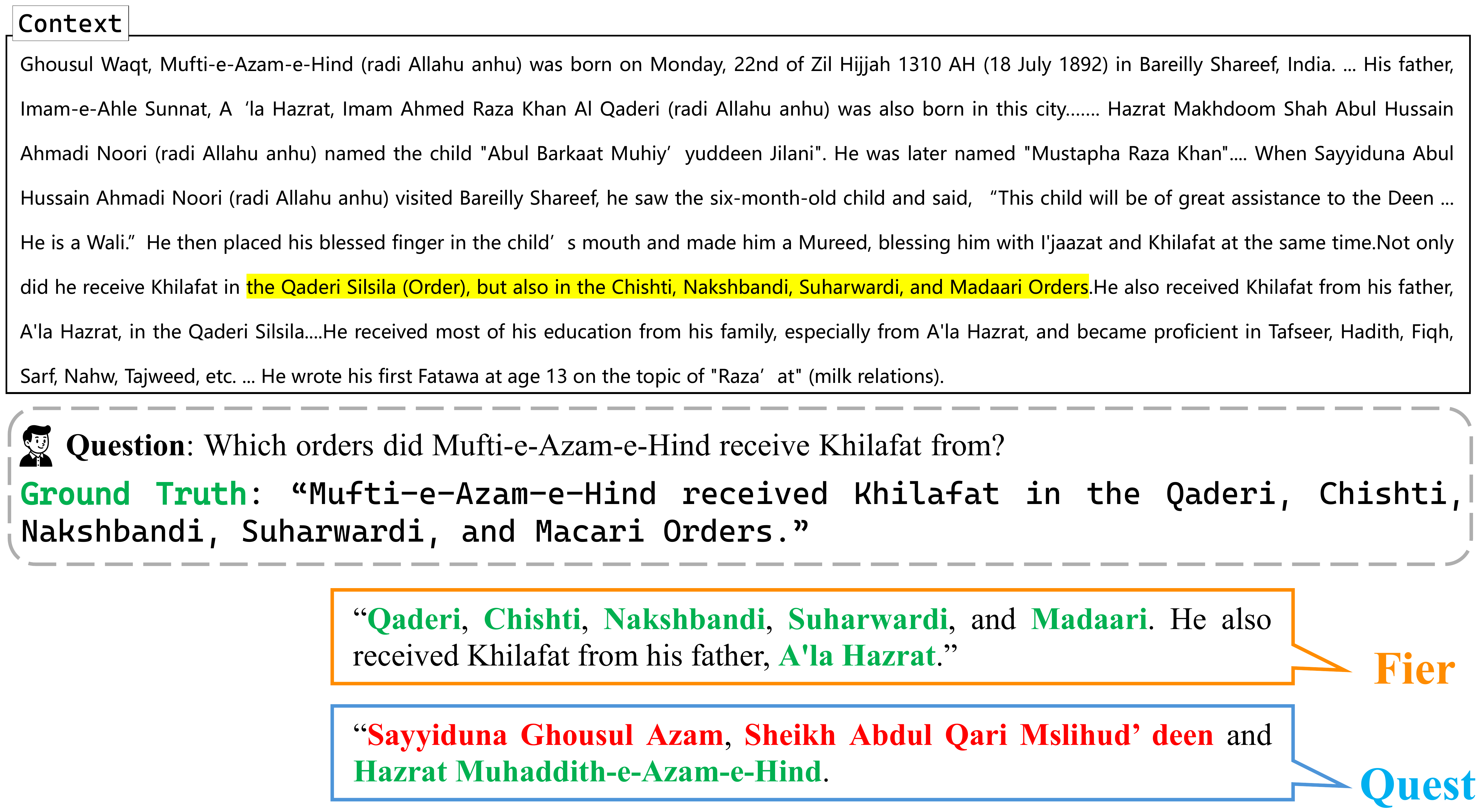}
    \includegraphics[width=.7\linewidth]{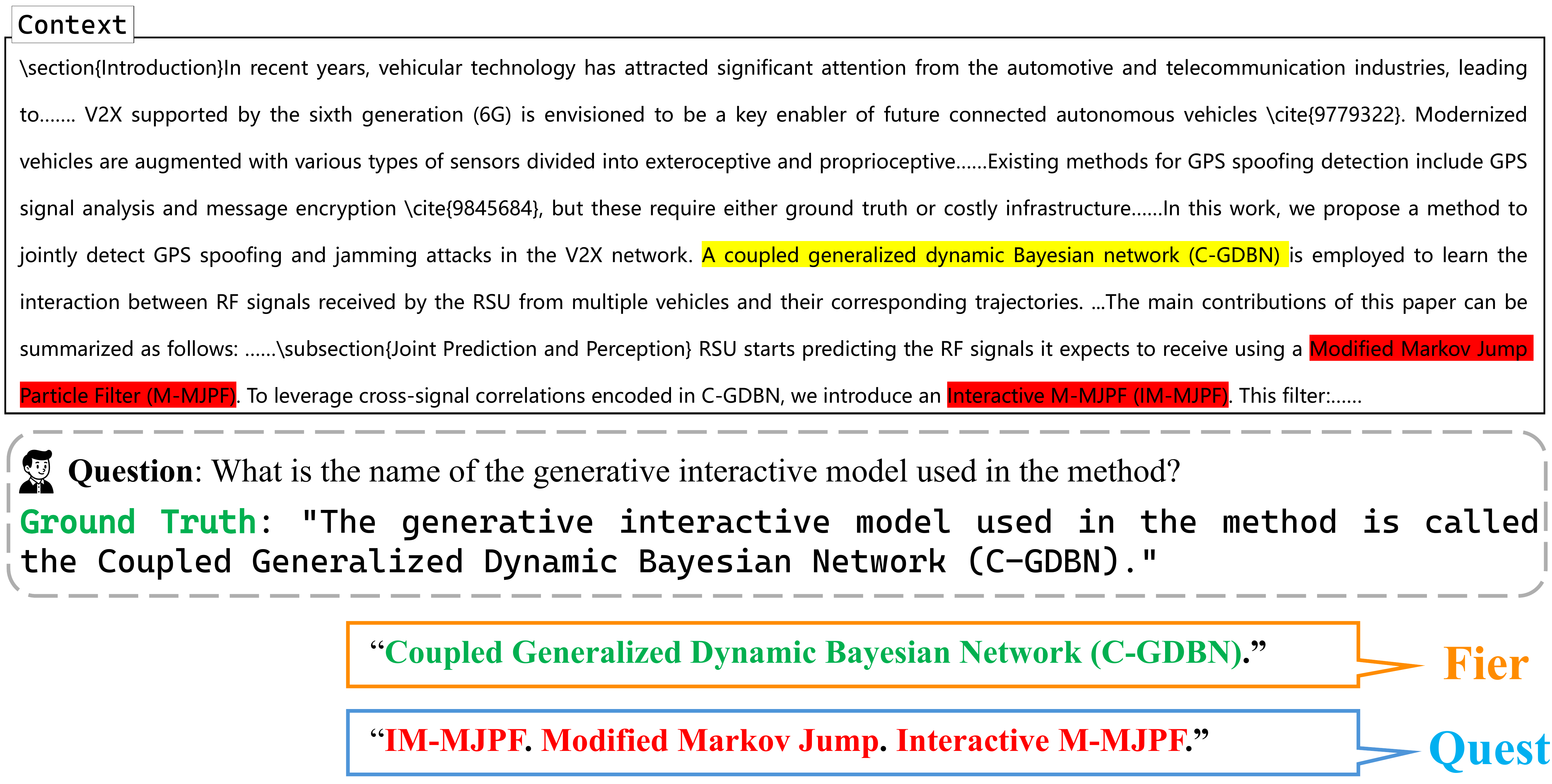}
    %\vspace{-10pt}
    \caption{Chatbot responses from Fier and Quest. Fier provides more complete and accurate answers in both examples.}
    \label{app1}
\end{figure*}
\label{AppendixC}
To better illustrate the practical differences between the two retrieval methods, we deploy a LLaMA-3-8B-Instruct chatbot using Fier and Quest, respectively. Given the same long context and user question, we present the corresponding responses from each chatbot for a qualitative comparison (Fig.~\ref{app1}).
 In the first example, which asks about the orders in which Mufti-e-Azam-e-Hind received Khilafat, the Fier-enabled chatbot correctly identifies all five orders, while the Quest-enabled chatbot only retrieves a single name, missing key information. A consistent trend is observed in the second example, which involves a scholarly article. When asked about the generative model adopted in the paper, the Fier-based chatbot accurately identifies the overall framework, whereas the Quest-based chatbot focuses narrowly on a sub-module mentioned in a later section.
\end{document}